\documentclass[journal]{IEEEtran}
\usepackage{cite}

\ifCLASSINFOpdf
  \usepackage[pdftex]{graphicx}
\else
 \usepackage[dvips]{graphicx}
\fi
\usepackage[cmex10]{amsmath}
\usepackage{url}

\begin{document}
%
\title{Development of a SQUID-based $^{3}$He Co-magnetometer Readout for a Neutron Electric Dipole Moment Experiment}
%
%
%

\author{Young Jin Kim and Steven M. Clayton
\thanks{Manuscript received October 9, 2012. This work was supported by the U.S.
Department of Energy under contract .}
\thanks{Young Jin Kim is with Los Alamos National Laboratory, Applied Modern
Physics Group, P.O. Box 1663, MS D454, Los Alamos, NM 87545, USA (email:
youngjin@lanl.gov).}
\thanks{Steven M. Clayton is with Los Alamos National Laboratory, Subatomic
Physics Group, MS H846, Los Alamos, NM 87545, USA (Corresponding
author: fax: +1 505 665 7920; e-mail: sclayton@lanl.gov).}
}

%
%

\markboth{2EF-08}
{Shell \MakeLowercase{\textit{et al.}}: Bare Demo of IEEEtran.cls for Journals}
%



\maketitle

\begin{abstract}
A discovery of a permanent electric dipole moment (EDM) of the neutron would provide
one of the most important low energy tests of the discrete symmetries beyond the
Standard Model of particle physics. A new search of neutron EDM, to be conducted at
the spallation neutron source (SNS) at ORNL, is designed to improve the present
experimental limit of $\sim 10^{-26}$e$\cdot$cm by two orders of magnitude. The
experiment is based on the magnetic-resonance technique in which polarized neutrons
precess at the Larmor frequency when placed in a static magnetic field; a non-zero
EDM would be evident as a difference in precession frequency when a strong external
electric field is applied parallel vs. anti-parallel to the magnetic field. In addition
to its role as neutron spin-analyzer via the spin-dependent n+$^{3}$He nuclear capture
reaction, polarized helium-3 (which has negligible EDM) will serve as co-magnetometer
to correct for drifts in the magnetic field.
In one of the two methods that will be built into the apparatus,
the helium-3 precession signal is read out by SQUID-based gradiometers.
We present a design study of a SQUID system suitable for the neutron EDM apparatus,
and discuss using very long leads between the pickup loop and the SQUID.
\end{abstract}


%
\IEEEpeerreviewmaketitle

\section{Introduction}
%
%
%
%
\IEEEPARstart{T}{he} search for a permanent electric dipole moment (EDM) of
elementary particles is
an
attempt to answer one of the most outstanding
questions in the understanding of fundamental physics and the physics universe~\cite{ramsey50}.
The existence of an EDM requires the violation of both the time-reversal (T) and
the parity-inversion (P) symmetries due to their different transformation properties
under discrete symmetry operations~\cite{landau57}.
Hence, the EDM search provides direct information
about the nature of T violation which is essential for explaining the observed baryon
asymmetry of the universe. The Standard Model (SM)
predicts a neutron EDM (nEDM) on the order of
10$^{-31}$e$\cdot$cm~\cite{cpv_without_strangeness},
well below the current experimental limit of
10$^{-26}$e$\cdot$cm~\cite{baker06}, while many theories of physics beyond the SM lead
to a sizable nEDM. The EDM search presents a powerful tool for tests of theoretical
extensions to the SM~\cite{cirigliano10}.

In the attempt to improve the present experimental limit by two orders of magnitude,
a new experimental search of nEDM, to be conducted at the spallation neutron source (SNS)
at Oak Ridge National Laboratory, has been proposed~\cite{nedm_proposal}.
In this paper, we describe the new
nEDM experiment and focus on efforts to design a SQUID system into the
nEDM cryostat for helium-3 magnetometer readout.

\section{Experimental Details}
\subsection{Background}
The technique to determine the nEDM, $d_n$, is based on magnetic-resonance.
Polarized neutrons precess at the Larmor frequency when placed in a static magnetic
field, and a non-zero EDM would induce a shift in the precession frequency when the
neutrons are subject to a strong external uniform electric field applied parallel
or antiparallel to the magnetic field.
The frequency shift due to an EDM is given by
\begin{equation}
\Delta\omega = -\frac{2 d_n E }{\hbar},
\end{equation}
where $E$ is the strength of the electric field applied along the magnetic field.
Using the experimental goal precision
of $\sim$10$^{-28}$e$\cdot$cm as the size of the nEDM and an electric field of 50 kV/cm,
the EDM-induced frequency shift is $\sim$2~nHz, equivalent to a precession
frequency shift due to a change in the magnetic field of $\sim$0.1~fT.
Thus, an EDM measurement to this level of precision requires extremely accurate
monitoring of the magnetic field that the neutrons experience.
Also required are a strong electric field, long observation times, and a
large number of neutrons for statistical precision.


\subsection{Experimental Method}

The experimental concept, due to Golub and Lamoreaux~\cite{golub94},
employs ultracold neutrons (UCNs) --- very low kinetic energy neutrons that
can be confined to a material trap~\cite{golubUCN} ---
and $^3$He atoms dissolved in superfluid $^{4}$He.
The $^3$He has a negligible EDM
due to Schiff screening~\cite{schiff63} of the nucleus
and serves as a magnetometer over the same volume occupied by the UCNs,
a so-called co-magnetometer.
The $^3$He acts as neutron spin-analyzer via the spin-dependent nuclear capture reaction
$\textrm{n} + {}^3\textrm{He} \rightarrow {}^3\textrm{H} + \textrm{p} + 764~$keV,
which proceeds with a rate proportional to $(1 - \cos{\phi_{n3}})$, where
$\phi_{n3}$ is the angle between the spin directions of the neutron
and $^3$He.
The triton and proton recoil products scintillate in the helium-4, and these
flashes of light are detected.

To extract the nEDM signal, the apparatus will be capable of two different
methods that we refer to as
``dressed spin'' and ``free precession'' modes.
In the dressed spin mode, a relatively strong, high frequency alternating magnetic field
(``dressing field'') is applied transverse to the static magnetic field $B_0$.
The parameters of the dressing field are chosen such that the neutron and $^3$He acquire
the same effective gyromagnetic ratios.  A modulation-feedback scheme, acting on the
dressing field parameters and based on the observed $\textrm{n} + {}^3\textrm{He}$
capture reaction scintillation signal, measures the nEDM (see Ref.~\cite{golub94}
for an extensive discussion of this technique).

The free precession mode is conceptually simpler and will be treated
in the remainder of this paper: no dressing field is applied,
and the UCNs and $^3$He are allowed to precess in a static magnetic field $B_0$
and parallel or anti-parallel electric field $E$ for a given measurement period,
while scintillation events are recorded ($^3$He as neutron spin-analyzer)
and the $^3$He magnetization signal is directly detected ($^3$He as magnetometer).
The scintillation rate $R_\gamma(t)$ from neutron capture on $^3$He will be modulated as the
polarization vectors of the neutron and $^3$He ensembles come into and out of alignment,
\begin{equation}
R_\gamma(t) \propto 1-P_nP_3\cos(2\pi\nu_s t+\varphi)
\end{equation}
where $P_n$ and $P_3$ are polarization vectors,
$\nu_s = \nu_3 - \nu_n$ is the difference
between the $^3$He ($\nu_3$) and neutron ($\nu_n$) precession frequencies,
and $\varphi$ is a constant phase.
The $^3$He and neutron gyromagnetic ratios $\gamma_n$, $\gamma_3$ differ
by about 10\%.  Hence, in a static, uniform
magnetic field, the scintillation event rate is modulated at about 10\% of the $^3$He
precession frequency. After correcting for any changes in the magnetic field,
a difference in the scintillation modulation frequency, $\nu_s$,
when the electric field is reversed would indicate
a non-zero nEDM signal. The nEDM, therefore, is extracted by the equation:
\begin{equation}
d_n=\frac{\hbar}{2E}\bigg[2\pi(\nu_{s}^{\uparrow\uparrow}-\nu_{s}^{\uparrow\downarrow})-\frac{(\gamma_3-\gamma_n)}{\gamma_3}2\pi(\nu_3^{\uparrow\uparrow}-\nu_3^{\uparrow\downarrow})\bigg],
\end{equation}
where the symbols $\uparrow\uparrow$ ($\uparrow\downarrow$)
refer to the electric field parallel (antiparallel) to the magnetic field.
The EDM sensitivity is estimated via error propagation to be
\begin{equation}
\begin{split}
\delta d_n = \frac{\pi\hbar}{E}\bigg\{ & (\delta \nu_{s}^{\uparrow\uparrow})^2
+(\delta \nu_{s}^{\uparrow\downarrow})^2 \\
   &  +\frac{(\gamma_3-\gamma_n)^2}{\gamma_3^2}\left[(\delta \nu_3^{\uparrow\uparrow})^{2}
+(\delta \nu_3^{\uparrow\downarrow})^2\right]\bigg\}^{1/2}.
\end{split}
\end{equation}
We note that the closeness of the gyromagnetic ratios of the two species
relaxes the measurement precision requirement on $\nu_3$ by a factor of 10.

The $^3$He precession signal will be read out with SQUID gradiometers.
An estimate of the resolution of the $\nu_3$ measurement, in the limit that
transverse relaxation time is much longer than the measurement period,
is~\cite{chibane95}
\begin{equation}\label{eq_chibane}
(\delta \nu_3)^2=\bigg(\frac{\delta\Phi_{sq}}{\Phi_{sq}}\bigg)^2\frac{3}{\pi^2t_m^3},
\end{equation}
where $\Phi_{sq}$ is the signal flux in the SQUID loop, $\delta\Phi_{sq}$ is the flux noise
spectral density in the SQUID at the signal frequency, and $t_m$ is the measurement
period. Taking into account the expected statistical precision of the $\nu_s$ measurement,
the goal of the co-magnetometer readout is $\delta \nu_3\ll 26\ \mu$Hz per 800 second
measurement period, such that $\delta\nu_3$ does not contribute significantly to
the overall uncertainty of the experimental result.

\subsection{Experimental Parameters}

\begin{figure}[!t]
\centering
\includegraphics[width=3.5in]{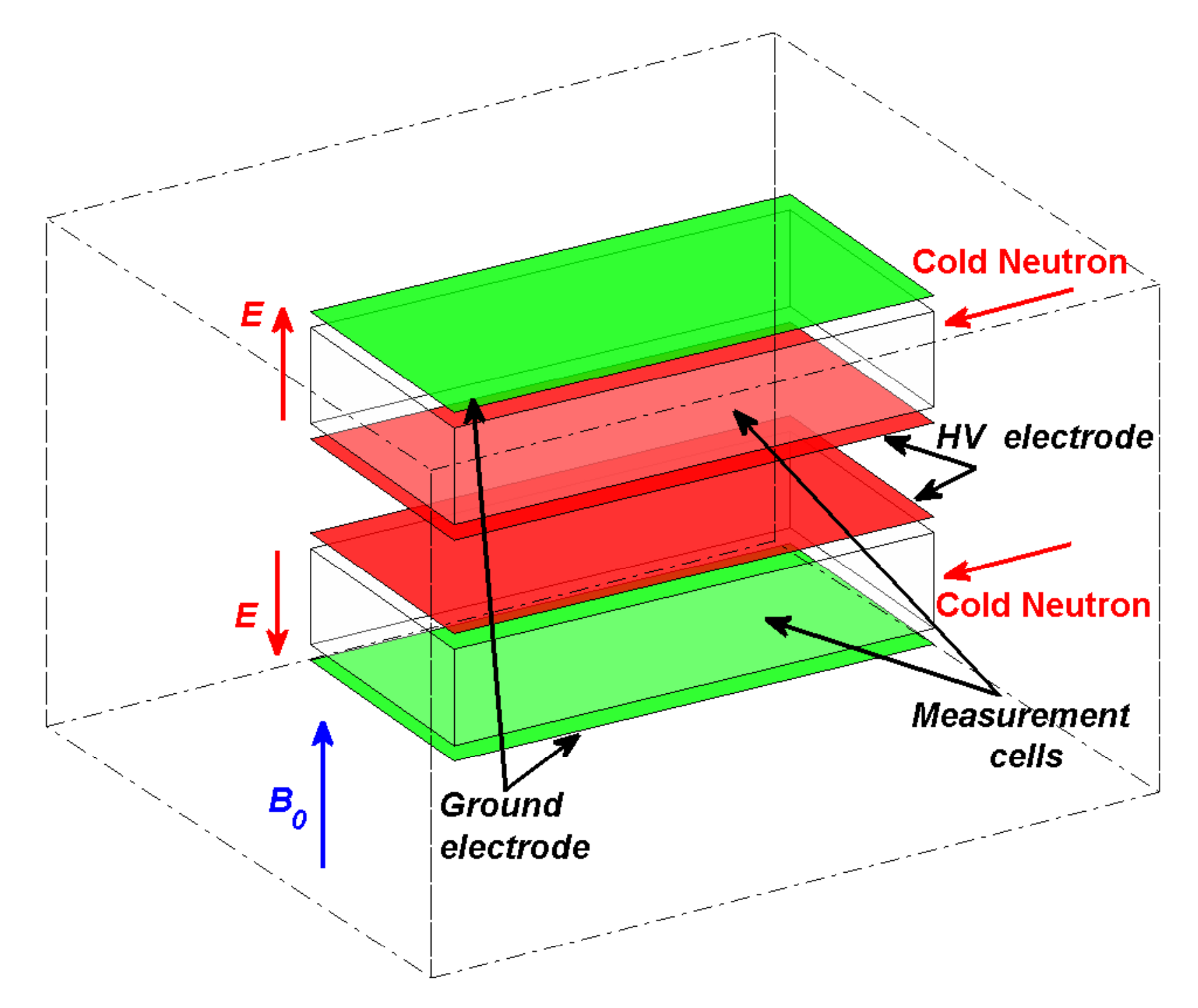}
\caption{Conceptual drawing of the nEDM apparatus.}
\label{cv}
\end{figure}

The choice of operating conditions --- cell temperature and geometry,
magnetic field, $^3$He density, etc. --- is driven by the statistics
requirement of the neutron spin-analyzer events
(${\rm n + {}^3He}$ captures) that measure the beat frequency $\nu_{n3}$,
and by known systematic errors that must be minimized.
Here, we will describe the conditions, in which the co-magnetometer readout
must operate, and leave further explanations to Refs.~\cite{nedm_proposal}
and~\cite{golub94}.
Two independent, identical, rectangular ($7.6\times10\times40$~cm$^3$)
measurement cells are arranged symmetrically
on either side of a high-voltage ($\approx$500~MV) electrode with ground electrodes on
the outboard sides of the cells, as shown in Fig.~\ref{cv}.
An applied, highly uniform magnetic field $B_0 \approx 3\ \mu$T is common to both cells;
thus, the $E$ and $B$ fields are parallel in one measurement
cell and antiparallel in the other.
The cells are filled with isotopically purified helium-4 at
nominal temperature $T_m = 450$~mK (which may changed to as low as $\approx$300~mK
for systematics studies) and contained in
a 1000-liter, helium-4-filled G10/composite vessel
cooled to the operating temperature by a dilution refrigerator.
Approximately 100\%-polarized $^3$He from an atomic beam source are filled into
both cells to a number density $\approx 2\times 10^{12}$/cm$^3$.
UCNs are produced in the cells \emph{in situ}, starting with a cold neutron
beam from the ORNL Spallation Neutron Source, by a phonon recoil process with
the superfluid helium-4; up to $\sim$100~UCN/cc are expected.

\begin{figure}[!t]
\centering
\includegraphics[width=3.5 in]{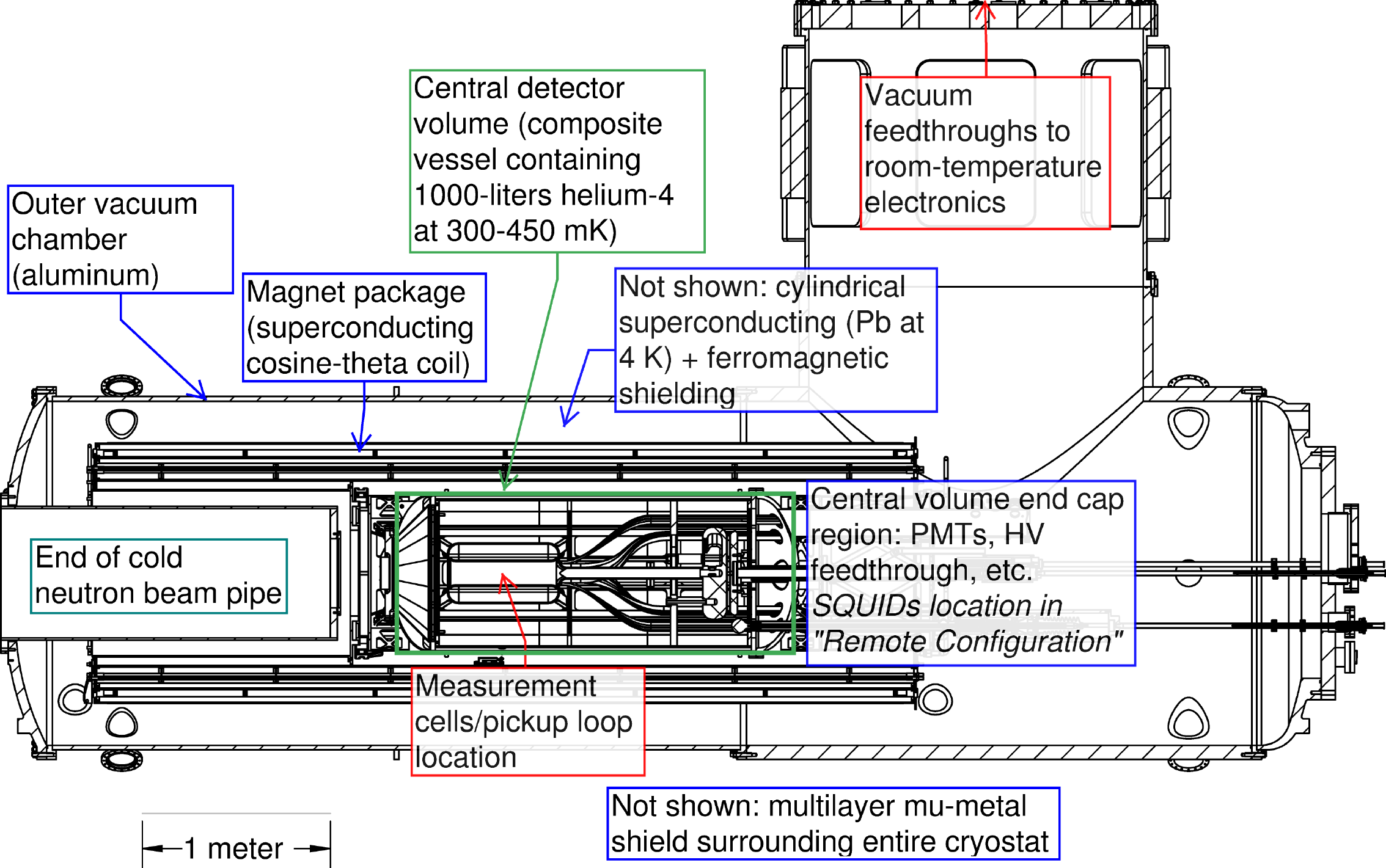}
\caption{Simplified and annotated cross-sectional drawing of the nEDM cryostat.}
\label{apparatus}
\end{figure}

The infrastructure to create these experimental conditions and perform
the measurement cycles is considerable.
Research and development,
and
preliminary engineering, have led to the apparatus depicted in Fig.~\ref{apparatus}.


\subsection{Magnetic Field}

Even with the co-magnetometer, control of the magnetic field environment
is critically important.
The holding field $B_0$ will be produced with an optimized cosine-theta
magnet coil with superconducting windings cooled to 4~K.
Surrounding the $B_0$ coil is a cylindrical ferromagnetic shield that
improves the uniformity of the $B_0$ field over the measurement cells.
Proceeding radially outward, next is a cylindrical superconducting shield
(Pb at 4~K), then the aluminum vacuum vessel of the cryostat, and
finally a multilayer high-permeability shield surrounding the entire cryostat.

Known systematic effects in this experiment are caused by magnetic field
non-uniformities in the measurement cells.
The ``$\vec{v}\times \vec{E}$ false-EDM'' effect,
a frequency shift linear in $\vec{E}$ that mimics an EDM,
is caused by an interplay of motional $\vec{v}\times \vec{E}$
fields (where $\vec{v}$ is an individual particle velocity) with
gradients in the static magnetic field~\cite{barabanov06}.
This and other known systematic effects sets the magnetic
field uniformity requirement of 
roughly $\langle\partial B_x/\partial x\rangle_V\leq10$~fT/cm, where the brackets
with subscript $V$ indicate volume averaging over the measurement cell.
The amount and geometry of magnetic and superconducting materials near
the cells must be restricted, since they could distort the $B_0$ field.

\section{Helium-3 Co-magnetometer Readout}

In the free-precession version of the nEDM experiment, the $^3$He magnetization
signal will be directly detected with superconducting gradiometers coupled to SQUIDs.
With $B_0 = 3\ \mu$T, the signal frequency is approximately 100~Hz.
Applying Eq.~\ref{eq_chibane} with measurement time $t_m = 800$~seconds and
requiring $\delta\nu_3 \ll 26\ \mu$Hz, we find the signal-to-noise
ratio in the SQUID must satisfy $\Phi_{sq}/\delta\Phi_{sq} \gg 1~\sqrt{\rm Hz}$.
The available signal is quite small because of the very low helium-3
concentration.  Furthermore, the pickup loops must be positioned
behind the ground electrodes of each cell to avoid high-electric-field regions.
The feasibility of SQUID detection of the $^3$He precession frequency
in these conditions has been demonstrated (by extrapolation)
with a simple setup~\cite{savukov08}.
However, the nEDM apparatus presents technical challenges to full
implementation of a SQUID-based co-magnetometer readout, such as
1) the long distance ($\sim$7~meters) between the measurement cells and room-temperature
electronics, 2) limitations on shielding such as superconducting conduit near the
measurement cells, 3) other devices (photomultiplier tubes, temperature sensors),
which could potentially emit RF inteference, must operate during the measurement period.

\subsection{Pickup Loops}

We propose a thin-film planar gradiometer as a pickup loop consisting of two
series-configured 3~cm$\times$6~cm pickup loops with center-to-center spacing of 9~cm.
The pickup loop is fabricated on a 150~mm Si wafer, and the inductance
is estimated to be $\approx$540~nH.
Eight individual gradiometers behind each ground plane and distributed across
the cell face,
with a nominal distance of 3~cm between the nearest part of the gradiometer and the inside
of the cell, are considered.

\begin{figure}[!t]
\centering
\includegraphics[width=3.3in]{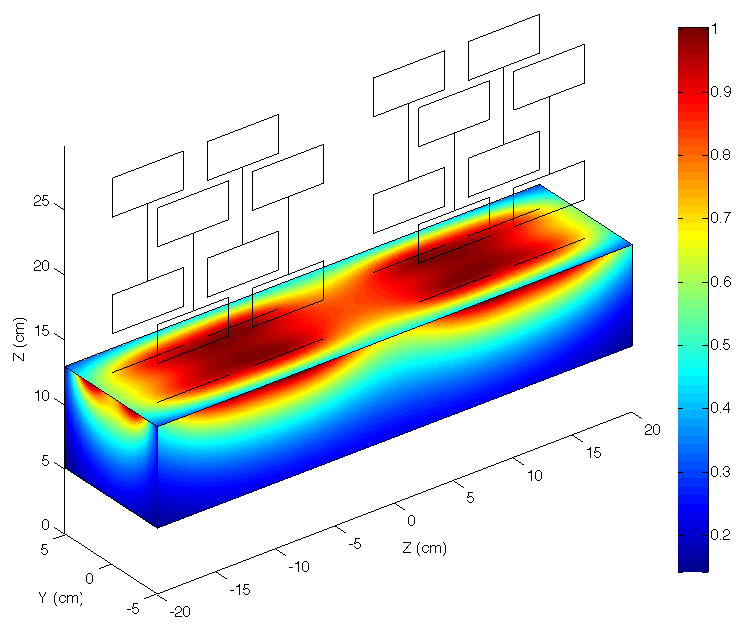}
\caption{Relative sensitivity of the gradiometer array vs. source voxel position,
for pickup loops separated by 3~cm from the inside of the nearest cell wall,
with source magnetization directed in the $+y$ direction.  In the coordinate
system shown, $\vec{B}_0$ is along the $x$-axis.}
\label{sensitivity}
\end{figure}

We calculate the signal flux amplitude through the array of eight gradiometers
for one EDM cell with $^3$He concentration $2\times10^{12}$~atoms/cm$^3$.
The magnetic flux through
each pickup loop due to a $^3$He dipole aligned with the $y$-axis and located
at $\vec{r}$ is computed. Then, numerical integration of $\vec{r}$
over the cell volume, appropriately scaled with the helium-3 number density, gives the
total flux expected in the pickup loops. The average flux through each pickup (total
through the gradiometer array divided by number of pickups) is calculated to
be 2000~$\mu\Phi_0$, where $\Phi_0$ is the magnetic flux quantum;
dividing by the half-gradiometer area 18~cm$^2$ gives an equivalent
average signal amplitude of 2.3~fT.
The relative sensitivity of the gradiometer array to voxel position is shown in
Fig.~\ref{sensitivity}. We see that the coverage is good across the ground-side
face of the cell but falls off fairly quickly for sources deeper in the cell.
However, because of the large diffusion constant of
dilute $^3$He in $^4$He at the measurement temperatures
($\approx$400~cm$^2$/s at 0.45~K~\cite{diffusion}), each spin traverses the cell many times
during the measurement period, and thereby samples the whole cell volume.

\subsection{SQUID Configuration}

The signal in the pickup loop must be coupled into a SQUID for detection.
The flux in the SQUID $\Phi_{sq}$ is related to the flux in
the pickup $\Phi_e$ by
\begin{equation}\label{eq_phisq}
\Phi_{sq}=\frac{M}{L_p+L_{par}+L_i}\Phi_e=\alpha\Phi_e,
\end{equation}
where $M$ is the SQUID input loop coupling, $L_i$ is the SQUID
input loop inductance, $L_p$ is the pickup loop inductance, 
and $L_{par}$ is parasitic inductance in the leads connecting
the pickup loop to the SQUID input loop.

We consider two SQUID configurations with the gradiometer pickup loops.
One we term the ``Piggybacked SQUID,'' in which the SQUID mounted atop the gradiometer
pickup and connected with very short leads. The other is the ``Remote SQUID'' configuration
in which the SQUID is mounted outside the Central Volume and connected with
long ($\approx$3.5~m) superconducting twisted-pair leads.

The advantage of the Piggybacked SQUID configuration is to maximize the possible
signal-to-noise ratio (SNR) by minimizing parasitic inductance. On the other hand, more wires
(six or eight per SQUID) to control the SQUID outside the nEDM cryostat must be inside
the Central Detector volume; it would be much more difficult to replace the SQUID if damaged
because the Central Detector volume would have to be opened; and
heating the SQUID to expel trapped flux may be difficult (though perhaps possible
with suitable insulation) since it is immersed in the 1000-liter superfluid helium-4 bath.

In the case of the Remote SQUID configuration, there are more
advantages in spite of the reduction of SNR due to the large parasitic inductance
of the twisted-pair leads (we measured 2.3~nH/cm for 3-mil Nb wires):
(a) only two wires per SQUID would be in the Central volume;
(b) minimal superconducting material near the measurement cells;
(c) replacement of SQUIDs is easier;
(d) the SQUID itself can be housed in a superconducting enclosure.
We are presently pursuing this configuration because of these advantages
and indications that the SNR will be sufficient.

For the Remote SQUID configuration, a high-inductance SQUID (StarCryo SQ2600, with
$L_i$ of 2585~nH and $M$ of 21~nH) has been chosen to increase the SNR because the
parasitic inductance is relatively less important.
We measured the intrinsic noise in the SQUID sensor with
a dummy load of 2600~nH to be 3.4~$\mu\Phi_0/\sqrt{\text{Hz}}$ at 1 kHz and 4~Kelvin.
Furthermore, we tested the flux noise
with a 3.5~m niobium twisted-pair lead connected to the input coil. The far end of the
leads was shorted. We confirmed that the noise is roughly the same as with the above,
which means long leads do not degrade the SQUID's performance.
We find the optimal number $N$ of pickup loop turns
by scaling the signal flux by $N$ and the pickup loop inductance by $N^2$
in Eq.~\ref{eq_phisq}, with $L_{par} = 805$~nH for the 3.5~m leads.
Two or three turns give the same, optimal SNR of $4.6~\sqrt{\text{Hz}}$,
satisfying the nEDM requirement.
Owing to eight gradiometers, the SNR may be further increased by a factor
of $\sqrt{8}$. In addition, the SQUID intrinsic noise may be improved 
at the operating temperature, $T_m$, much
lower than 4.2~K as much as a factor of $\sqrt{4.2 {\rm K}/T_m }$~\cite{espy02}.

\section{Future Work}

We plan to study experimentally the Remote SQUID configuration
with shielding and magnetic fields similar to the planned nEDM apparatus.
Furthermore, we will test the compatibility of low-noise
SQUID operation with other devices that are potential
sources of electromagnetic interference, especially photomultiplier tubes,
which are necessarily operating during the measurement period.

\section{CONCLUSION}
A new search for nEDM aiming to 100-fold improvement in the present experimental
limit is described. This experiment employs polarized helium-3 as neutron spin-analyzer
and co-magnetometer. In one operating mode of the experiment, the $^3$He co-magnetometer
is read out by first-order planar gradiometers.
A configuration in which the gradiometer pickup loop
is attached to the SQUID by very long leads appears feasible and
offers some advantages over a piggybacked configuration.

\section*{Acknowledgment}
We thank Robin Cantor, Michelle Espy and Andrei V. Matlashov for assistance
with SQUID system tests as well as very helpful discussions. 
This work was supported by the US~DOE Office of Science, Nuclear Physics.

\ifCLASSOPTIONcaptionsoff
  \newpage
\fi



%

\bibliographystyle{IEEEtran}


%




\end{document}